\documentclass[journal=jpcb,manuscript=article]{achemso}
\usepackage{chemformula} 
\usepackage[T1]{fontenc} 
\usepackage{float}
\usepackage{subfig}
\usepackage{soul}
\usepackage{hyperref}
\usepackage[version=3]{mhchem} 

\author{Clifford E. Woodward}
\affiliation{School of Physical, Environmental and Mathematical Sciences University College, University of New South Wales, ADFA Canberra ACT 2600, Australia}
\author{David Ribar}
\affiliation{Computational Chemistry, Lund University, P.O.Box 124, S-221 00 Lund, Sweden}
\author{Jan Forsman}
\email{jan.forsman@compchem.lu.se}
\affiliation{Computational Chemistry, Lund University, P.O.Box 124, S-221 00 Lund, Sweden}
\title{Classical density functional treatment of polydisperse polarisable clusters}
\keywords{polyampholytes, electrolytes, ion clusters, simulations, anomalous screening, colloidal stability}

\begin{document}
\begin{abstract}
  Ion clustering has been proposed as a mechanism leading to the peculiar ``anomalous underscreening'' phenomenon
  seen for electrostatic interactions between charge surfaces immersed in concentrated electrolytes.
  These interactions have been measured using the Surface Force Apparatus, according to which there are
  strong repulsive interactions between like-charged surfaces,
  with a range that increases upon further addition of salt,above some threshold concentration.
  A common suggestion is that ionic aggregates, if they form in sufficient numbers,  will reduce the
  concentration of free ions and thereby increase the nominal Debye length.
  In previous work, we investigated a cluster model using classical Density Functional Theory (cDFT)
  and a polymer-like description of the ion clusters.  These clusters were monodisperse and of either a linear or branched
  architecture, and a fixed charge sequence along the chains.
  In this work, we generalise the cDFT to treat ``living polymers'' with
  variable chain lengths and charge arrangements along the chain. 
  This approach allows clusters to become polarised by the presence of charged surfaces,
  manifested by like-charged bonding.  We find that even with a small degree of like-charged bonding 
  a full equilibrium treatment of our model predicts only weak repulsion between like-charged surfaces.
  When a global constraint is applied so that the charged surfaces are neutralised only by the dissociated ions, while the clusters 
  contribute overall zero charge, even a very small fraction of clustering ions generate strong and long-ranged forces. Moreover, if
  the cluster fraction increase substantially upon the addition of further salt, then the 
  strength of the surface forces will also increase, although the range remains roughly constant.
    \end{abstract}

\section{Introduction}
Interactions between charged surfaces immersed in electrolyte solutions have been the subject of extensive study for
decades, owing to their fundamental importance in numerous biological and industrial systems. A widely used theoretical
framework is based on the primitive model, where the solvent is treated as a dielectric continuum, and ion distributions
are described within the mean-field Poisson–Boltzmann (PB) theory. For instance, this model forms the
basis of the seminal DLVO theory, describing colloidal interactions \cite{Derjaguin:41,Verwey:48,Israelachvili:91}.

Extensions to the PB approach that account for ion–ion correlations are essential for
understanding systems characterised by strong electrostatic coupling. Related phenomena, such as charge inversion and like-charge
attraction, are now relatively well-understood.
Within mean-field treatments of primitive models, the characteristic length scale governing electrostatic screening is the
Debye length, $\lambda_D$. This length quantifies the effective range of electrostatic interactions between charged
surfaces in dilute electrolytes and decreases with increasing ionic concentration and/or charge, reflecting enhanced screening. However, recent
experimental observations, primarily using the Surface Force Apparatus (SFA), have revealed striking deviations from this classical
behaviour \cite{Gebbie:13, Gebbie:15, Smith:16, Fung:23, Elliot:24}. These studies report that the decay length of surface interactions
in simple aqueous electrolytes and in ionic liquids can substantially exceed $\lambda_D$, and, intriguingly, may increase
with salt concentration beyond a certain threshold concentration. This counter-intuitive result is often referred to as
{\em anomalous underscreening} \cite{Hartel:23}.

Independent experimental evidence for anomalous underscreening has been obtained using techniques other than SFA.
For example, Gaddam and Ducker \cite{Gaddam:19} employed fluorescence imaging to probe the spatial distribution of
a negatively charged fluorescent tracer (at micromolar concentration) between charged silica surfaces. Their
measurements revealed concentration-dependent decay lengths consistent with the anomalous trends seen in SFA experiments. Such
an inverse relationship between screening length and ionic strength would also be expected to influence colloidal
stability. Indeed, Yuan et al. \cite{Yuan:22} reported re-entrant dispersion behaviour in charged colloids, where
aggregation at intermediate salt concentrations was followed by re-dispersion at very high concentrations.

It should also be noted that Kumar {\em et al.} \cite{Kumar:22} employed atomic force microscopy (AFM) to study
interactions between charged surfaces in aqueous electrolytes, but did not observe anomalous underscreening; at high
ionic strengths, the measured forces were short-ranged. Baimpos {\em et al.} investigated $LiCl(aq)$ and $CsCl(aq)$ solutions
using both SFA and AFM, finding long-range forces with the former but not the latter technique. They suggested that SFA
measurements more accurately capture equilibrium interactions, whereas AFM may be limited by a lower sensitivity.

Although anomalous underscreening has now been reported by several independent groups, its physical origin remains
unresolved \cite{Lee:17,Coupette:18,Rotenberg:18,Kjellander:20,Coles:20,Zeman:20,Cats:21,Krucker:21,Hartel:23,Elliot:24,Ribar:24}.
Existing theoretical models do not yet provide a convincing explanation. A commonly proposed mechanism
involves ionic clustering \cite{Gebbie:13,Gebbie:15,Ma:15,Hartel:23,Ribar:24,Ribar:25a,Ribar:25b}, which is
expected to become increasingly significant at high concentrations and strong
coupling. In this view, the observed screening length reflects a renormalised Debye length determined primarily
by a reduced population of “free” ions, since ion clusters are likely to carry little or no net charge.

Experimental evidence for ion clustering at high ionic strengths has also been
reported \cite{Sedlak:06a,Sedlak:06b, Liao2025, Irving2023, Georgalis2000, Fetisov2020, Shalit2016, Straub2022}. 
For example, concentrated $KCl(aq)$ and $NaCl(aq)$ solutions
exhibit fluorescence \cite{Villa:22}, which Villa {\em et al.} have attributed to a stiffening of hydrogen-bond networks
in hydration shells, thereby suppressing non-radiative decay pathways. Their interpretation was supported by
quantum chemical calculations, suggesting that the fluorescing structures correspond to hydrated ion clusters.

However, ion clustering may in principle affect surface interactions in more ways than merely via the
concomitant drop of the ionic strength. A recent study utilising classical polymer Density Functional Theory, cDFT, suggested
that solutions in which {\em all} ions are clustered, and treated as polymers, mediate remarkably strong and long-ranged repulsive
forces between charged surfaces \cite{Ribar:25a}. There are a few noteworthy limitations to the model used in
that work:
\begin{itemize}
 \item the polymers formed (``ion clusters'') were overall monovalent and monodisperse
 \item the polymers had a fixed sequence of charges (in most cases alternating)
 \item there were no dissociated ions (simple salt) present
\end{itemize}
The model used a linear polymers to model the clusters.  This might also have seemed crude, but essentially identical
results were established with corresponding models utilising branched architectures.
This implies that the linear sequence constitutes a reasonable model (note that even linear polymers will typically have a rather
globular shape in solution).

In this work, we will generalise the cDFT to handle 
mixtures containing two different kinds of
monomers ($A$ and $B$, say), which in turn can polymerise to form {\em living}  (or {\em equilibrium})  polymers,
by forming reversible bonds ($A-A$, $A-B$ and $B-B$), which produce range of
chain lengths {\em and} compositions. Moreover, both of these
will respond to changes of the external conditions. For instance, the presence of a surface
with a strong affinity to $A$ will induce the formation of longer $A$-rich regions with the adjacent polymers.
As an application to the electrolyte systems discussed above, the monomers ($A$ and $B$) will refer to anions and 
cations specifically. The formation of like-charged bonded monomers in the chains ($A-A$ and $B-B$) while energetically
unfavourable (due to the Coulomb repulsion) will occur as a response to the application of external potentials due, e.g., to 
charged surfaces.  If such bonding was disallowed then chains would be restricted to a sequence of alternating charges,
which reduces the ability of our cluster model to manifest  charge asymmetry or become charged in general.  
Below, we will use the term "polarisation" to describe both these types of perturbation (asymmetry and charging) of the clusters.   
In the next section, we describe the cDFT in more detail.


\section{Model and Theory}
The cDFT formulation presented here is able to treat general $A/B$ mixtures of monomers 
that form linear polymers with "dissociable bonds'' of type $A-A$, $A-B$ and $B-B$.
In this application we are mainly concerned with the case where ``$A$'' is a
monovalent cation (say) and ``$B$'' is a monovalent anion, both of
which are dissolved in an implicit aqueous medium. There are special
aspects to consider for such systems, compared with neutral monomers $A$ and $B$.
For instance, their mutual Coulomb interaction is expected to render
$A-B$ bonds stronger than $A-A$ and $B-B$.  We note that like-charged
bonding may be advantageous in the presence of strongly charged surfaces of 
a valence which is opposite to that of the bonding monomers.  The mutual repulsion of like-charged 
monomers is counteracted by the added attraction to the surface.  Ion correlation 
effects can also enhance the presence of like-charged bonds as well as 
hydration forces. 

The general picture is as follows.  A local concentration fluctuation of ions may form large 
"bonded" clusters resembling a loose crystal, which traps some water
molecules and displaces others.  As already mentioned the degree of cluster polarisation
induced by the presence of charged surfaces is reflected in the degree  
of like-charged monomer bonding in our model.  The presence of localised regions of like-charge
in the cluster is also further facilitated by rearrangements of nearby ions and water molecules.
For this reason, in our model, we let the degree of like-charged and unlike-charged bonding
in the linear chain to be determined by variable parameters.


\subsection{Density Functional Theory of  linear polarisable ion clusters }
We begin with the exact canonical free energy density functional  for an ideal,
monodisperse polymer fluid of flexible $r$-mers, which can consist of different types of monomers, 
\begin{eqnarray}
\beta F^{id}_r & = & \int d{\bf R}  N_{\bf c}({\bf R}) 
\left( \ln [N_{\bf c}({\bf R})] - 1 \right) + \\ \nonumber
& & \int d{\bf R} N_{\bf c}({\bf R}) \Phi^{(b)}({\bf R}) + 
\int \sum_i d{\bf r} n_i({\bf r}) \psi^{0}_i ({\bf r}) 
\label{eq:idealb}
\end{eqnarray}
The vector subscript, ${\bf c} = (r, {\bf s})$, denotes both the degree of polymerization $r$ (chain length), 
and the sequence of monomer types, ${\bf s} = (s_1, s_2,...s_r)$, where the number of different monomer
types may be different to $r$.  In fact we now particularise to the case of  
only two different monomer types constituting "+" and "-" charges, so $s_i = +, -$.   

As usual, $\beta = 1 / (k_B T)$, denotes the inverse thermal energy.
The $r$-dimensional density, $N_{\bf c}({\bf R})$, is a function of the monomer
configuration ${\bf R} = ({\bf r}_1,...,{\bf r}_{r})$, where ${\bf r}_k$ is the
coordinate of monomer $k$ in the chain of type $s_k$ and $\Phi^{(b)}({\bf R})$ gives the
intra-molecular connectivity modelled as nearest-neighbour, non-directional bonding, i.e., 
\begin{equation}
\Phi^{(b)}({\bf R}) = \sum_{i=1}^{r-1} \phi^{(b)}(|{\bf r}_i-{\bf r}_{i+1}|)
\label{eq:NNbonding}
\end{equation}
We shall assume that the bonding term keeps bonded monomers at a constant distance $b$ apart.
Finally, $\psi^{(0)}_i ({\bf r})$, is the external potential acting on the monomer density, 
$n_i({\bf r})$,  of type $i = +, -$. 

The bulk fluid is assumed to contain equal numbers of "+" and "-" monomers in total, each with density $n_b$.  If the
the average density of aggregates is,  $\phi_p$, and $\langle r \rangle$ is the average association number, then $\phi_p\langle r \rangle = 2n_b$. 
The specific fractional distributions of polymer aggregate types is given by some function $F({\bf c})$.  This distribution gives the fraction
of species with internal structure ${\bf c}$ and is normalised according to
\begin{equation}
\sum_{\bf c} F({\bf c}) = 1
\end{equation}
The ideal chemical potential of the various polymer species, ${\bf c}$, in the bulk is then given by,
\begin{equation}
\beta \mu^{(id)}_{\bf c} = \ln[\phi_p F({\bf c})] 
\label{eq:chempot}
\end{equation}
In this study, we will assume that $F({\bf c})$ is such as to restrict polymers to consist of simple linear chains
of "+" and "-" monomers.  The chains have variable length as determined by
a fixed "bonding energy" between $NN$ monomers, which will allow segments of
chain where monomers of the {\em same charge} will be bound together.  In principle, such segments should 
be frustrated by like-charged repulsions.  To account for this we will assume that the distribution of
monomers is modified by $NN$ Coulomb interactions as well (see below). 

In the presence of additional (non-bonding) interactions, the ideal functional must be supplemented with a
suitable excess term, $F^{ex}$.  Thus the total Gibbs free energy functional is given by the following general expression,
\begin{equation}
\Omega =  F^{id}[\{N^{(c)}_{r}({\bf R})\}] + F^{ex}[\{n_i({\bf r})\}]-\sum_{r,c}\mu_{r,c}\int d{\bf R} N^{(c)}_{r}({\bf R}) 
\label{eq:grandpot}
\end{equation}
where $r$ represents the length of chain and $c$ is the specific distribution of "+" and "-" monomers.  More specifically
the chains will consist of alternating segments of like charges.  In the bulk fluid the average length of positive and negative
segments will be equal, but this will generally not be the case in the presence of an applied electrostatic potential. 

The first term is the total ideal contribution,
\begin{eqnarray}
\beta F^{id}[\{ N^{(c)}_{r}\} ] & = & \sum_{r,c}\int d{\bf R}  N^{(c)}_{r}({\bf R}) 
\left( \ln [N^{(c)}_{r}({\bf R})] - 1 \right) + \\ \nonumber
& & \sum_{r,c} \int d{\bf R} N^{(c)}_{r}({\bf R}) \Phi^{(b)}_{r}({\bf R}) + 
\sum _{i}\int d{\bf r} n_i({\bf r}) \psi_i^0 ({\bf r}) 
\label{eq:idealc}
\end{eqnarray}
The second term is the excess free energy, $F^{ex}[\{n_i({\bf r})\}]$,
 which depends upon the monomer densities, $n_+({\bf r})$ and $n_-({\bf r})$.  
The final term contains the chemical potential term given by,
\begin{equation}
\beta \mu_{r,{\bf s}} = \ln[\phi_p f(r,{\bf s})] + \beta \mu^{ex}_p(r)
\label{eq:chempotex}
\end{equation}
which consists of the ideal part $\ln[\phi_p f(r,\bf{s})]$.   Here
$\bf{s}$ is the vector of monomer types $\{ s(i) = +1, -1: i=1,r\}$. 
The excess term, $\mu^{ex}_p(r)$, is assumed to be independent 
of the charge of the polymer cluster in our theory, which is consistent
with the commonly used Restricted Primitive Model, RPM.
It will turn out (see below) that $\mu^{ex}_p(r)$ is a linear function of $r$.

Following the discussion above, the function $f(r,s)$ assumes
that the distribution of monomers in the polymer is determined by
a constant $NN$ bonding energy as well as $NN$ Coulombic interactions.  This gives,
\begin{eqnarray}
f(r, {\bf s}) & = & K e^{-\kappa_{++} N_L} e^{-\kappa_{+-} N_U} \\
\label{eq:Schultz-Flory}
\end{eqnarray}
where $N_L$ and $N_U$, are the number of like and unlike charged
bonded monomers,  respectively,  
\begin{eqnarray}
N_U & = & \sum_{i=1}^{r-1} | s(i) - s(i+1) |/2 \\
N_L & = & r- N_U - 1
\label{eq:config}
\end{eqnarray}
$K$ is the normalization factor and $\kappa_{++}$ and $\kappa_{+-}$ determines the degree of association.
We note that
$(\kappa_{++} - \kappa_{+-})/2 = \gamma$ is the $NN$ Coulombic interaction, which determines the 
difference in the like and unlike $NN$ pairs. 

In order to obtain $K$, we note that the number of monomers can also be written as $r$ = $\sum_{i= 1}^{N_{seg}} n_i$,
where $N_{seg} = N_U+1$ is the number of segments.  Adjacent segments have opposite charge and
their lengths are given by $\{ n_1, n_2....n_{N_{seg}}\}$. 
\begin{eqnarray}
K^{-1} & = & \sum_{N_{seg}=1}^{\infty} e^{-\kappa_{+-} (N_{seg}-1)} \prod_{i=1}^{N_{seg}}\sum_{n_i=1}^{\infty} e^{-\kappa_{++} (n_i-1)}  \\
 & = & \sum_{N_{seg}=1}^{\infty} \frac {e^{-\kappa_{+-} (N_{seg}-1)}} {(1- e^{-\kappa_{++}})^{N_{seg}}} \\
 & = & \frac{1}{ 1- e^{-\kappa_{++}} - e^{-\kappa_{+-}}}
 \label{eq:normalize}
\end{eqnarray}
It is instructive to obtain an expression for the total density of charged species in the bulk solution, $n_b = \phi_p \langle r \rangle $,
where $ \langle r \rangle$ is the average number of charged monomers per chain (we shall often use the more
compact expression ``chain length'' for this quantity).  Clearly the number
of positive and negative monomers per chain will be equal on average.  From the above expressions we obtain,
\begin{eqnarray}
n_b  & =  & \phi_p K \sum_{N_{seg}=1}^\infty e^{-\kappa_{+-} (N_{seg}-1)} 
\prod_{i=1}^{N_{seg}} \sum_{n_i=1}^{\infty} e^{-\kappa_{++}(n_i-1)}  \sum_{k=1}^{N_{seg}} n_k\\
& = & \phi_p K \sum_{N_{seg}=1}^\infty e^{-\kappa_{+-} (N_{seg}-1)} 
\prod_{i=1}^{N_{seg}} \sum_{n_i=1}^{\infty} e^{-\kappa_{++}(n_i-1)}  N_{seg}  \frac{\sum_{n=1}^{\infty} e^{-\kappa_{++}(n-1)}n}{\sum_{n=1}^{\infty} e^{-\kappa_{++}(n-1)}} \\
 & = & \phi_p \left< N_{seg}\right >  \left< n \right> 
\label{eq:nb}
\end{eqnarray}
where
\begin{eqnarray}
\left< N_{seg}\right >  & = & K \sum_{N_{seg}=1}^{\infty} e^{-\kappa_{+-} (N_{seg}-1)} \left[ \sum_{n=1}^{\infty} e^{-\kappa_{++} (n-1)}\right]^{N_{seg}}  N_{seg}\\
 \left< n \right> & = & \frac{ \sum_{n=1}^{\infty} e^{-\kappa_{++} (n-1)}n}{\sum_{n=1}^{\infty} e^{-\kappa_{++} (n-1)}}
\label{eq:avenseg1}
\end{eqnarray}
Thus  $\left< N_{seg}\right >$ is average number of segments per chain in the bulk and the average length of each segment is $\left< n\right >$.
The total density of charged monomers in the bulk can also be written as $n_b = \phi_p \left< r\right >$, where  $ \left< r\right > = \left< N_{seg}\right >\left< n\right >$. 
Now from Eq.(\ref{eq:normalize}) we obtain,
\begin{eqnarray}
\frac{\partial \ln K}{\partial \kappa_{++}} &  = & \left< N_{seg}\right > \left< n-1 \right> \\
& = & \frac {e^{-\kappa_{++}}}{1- e^{-\kappa_{++}} - e^{-\kappa_{+-} }}
\label{eq:average}
\end{eqnarray}
We can also write,  
\begin{eqnarray}
\left< N_{seg}\right > & = & \frac{\partial \ln K}{\partial \kappa_{+-}} + 1\\
 & = & \frac {1-e^{-\kappa_{++}}}{1- e^{-\kappa_{++}} - e^{-\kappa_{+-} }}
\label{eq:avenseg2}
\end{eqnarray}
These relations give, 
\begin{eqnarray}
\left< N_{seg}\right > \left< n \right >   & = & \frac {1}{1- e^{-\kappa_{++}} - e^{-\kappa_{+-} }}\\
& = & K^{-1}
\label{eq:avenseg3}
\end{eqnarray}
and thus, $<n> = [1-e{-\kappa_{++}}]^{-1}$.
\begin{figure}
	\centering
	\includegraphics[scale=1]{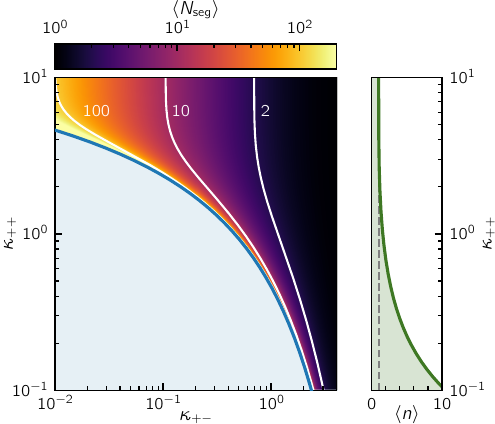}
	\caption{Visualisation of the $\langle N_{seg}\rangle, \langle n\rangle$ dependence on $\kappa_{++}$ and $\kappa_{+-}$. Regions
          with  $\kappa_{++} > \kappa_{+-}$ depict cases for which the backbone charges primarily (but not completely) 
          \textit{alternate} (in the absence of external fields), while $\kappa_{+-} > \kappa_{++}$ promotes
          the formation of \textit{block} charges.
          The blue line and shaded area indicate the unphysical region, where $1/K$ diverges, or becomes negative.
	}
	\label{fig:kappa_dep}
\end{figure}

The equilibrium polymer density is obtained by minimizing, $\Omega$, which gives the following 
self-consistent expression for the polymer density,
\begin{equation}
N_{r,\bf{s}}({\bf R})= \phi_P K \prod_{i=1}^{r} z_{s(i)}({\bf r}_i)\prod_{i=k}^{r-1}T(|{\bf r}_k-{\bf r}_{k+1}|)e^{-\kappa_{s(k)s(k+1)}}
\label{eq:rden}
\end{equation}
Recall, ${s(i)}$ =$\pm 1$, signifies the valency of the $i^{th}$ monomer.  The non-uniform fugacity term is given by
\begin{equation}
z_{s}({\bf r}) =  e^{-\beta \psi_{s} ({\bf r})}
\label{eq:nonunifug}
\end{equation}
Furthermore, we have defined,
\begin{equation}
\psi_{s}({\bf r}) =  \frac{\delta {F}^{ex}}{\delta n_s({\bf r})} -  \mu^{ex}_{s} + \psi_s^0({\bf r})  
\label{eq:lambda}
\end{equation}
Thus the potential $\psi_i({\bf r})$ contains the electrostatic potential on the monomer of type $i$ (= $\pm$),
which includes the Donnan potential, as well as an excluded volume contribution.  In addition,
we have a contribution from any non-electrostatic external potential.  We also subtract an excess chemical
potential term that is determined by the bulk conditions.  Referring to Eq.(\ref{eq:chempotex}), this implies 
that  
\begin{equation}
\mu^{ex}_{p}(r) = \sum_{i=1}^{r} \mu^{ex}_{i}
\label{eq:chempotex2}
\end{equation}
As stated earlier, in our simplified theory we will find that $\mu^{ex}_{i}$ will be independent 
of the monomer type $i$.  Finally, we have
\begin{equation}
T(|{\bf r}-{\bf r}'|) = \frac{e^{-\phi^{(b)}(|{\bf r}-{\bf r}'|)}}
{\int d{\bf r}'e^{-\phi^{(b)}(|{\bf r}-{\bf r}'|)}}
\end{equation}
We shall assume that monomers are bound at a fixed bond-length
$b$, irrespective of their type, though the bond-energy will depend upon the pair of
monomers involved, i.e., 
\begin{equation}
T(|{\bf r} -{\bf r}'|) = \frac{\delta (|{\bf r}-{\bf r}'|-b )}{4\pi b^2}
\end{equation}
It is convenient to define the following expressions: \newline
${g_{s}({\bf r}_i)} =  e^{-\beta \psi_{s} ({\bf r})/2}$\newline 
$S_{ss'}({\bf r}, {\bf r}') = g_{s}({\bf r})T(|{\bf r} -{\bf r}'|)g_{s'}({\bf r}')$  \newline
$F^{(s)}_{n}({\bf r}_1...{\bf r}_{n+1})=  \prod_{i=1}^{n} S_{ss}({\bf r}_i, {\bf r}_{i+1})$ \newline
$F^{(s)}_{0}({\bf r}_1) = {g_{s}({\bf r}_1)}$\newline
The last two terms corresponds to an $n$-segment distribution (unnormalised) of a single type of charged monomers (type $s$).  
We now note that the vector \{$s(1), s(2), ....s(r)$\} appearing in Eq.(\ref{eq:rden}) can be grouped into 
$n_i$-segments of like charged monomers, i.e., $N_{seg}$ segments of type \{ $s_i =\pm 1 : i = 1, N_{seg}$\}
with lengths given by  \{ $n_i : i = 1, N_{seg}$\}.  Note that the, $s_i$, will alternate between 
+1 and -1, and the sum of all the $n_i$ is equal to $r$. The total ensemble
of chains is obtained by summing over the number of segments, $N_{seg}$, and the number of (like-charged) monomers 
within each segment, $n_i$.  We reiterate that the segments will
alternate in charge, but for $N_{seg}$ odd, we will have the end segments be either both positive 
or both negative and, for $N_{seg}$ even, the end segments will be of opposite charge.    

Suppose for the moment we consider the subset of chains with a fixed number of segments, $N_{seg}$, but otherwise
all possible number, $n_i$, of like-charged monomers within each segment.  
Let us calculate the total positive charge density at some position, ${\bf r}$ for such chains.  We denote this density
as, $n^{(+)}_{N_{seg}}$.  We begin by defining the following contracted (unnormalised) segment density,
\begin{equation}
{\bar F}^{(s)}_{n}({\bf r}, {\bf r}')=  \int d{\bf r}_1... d{\bf r}_n \delta ({\bf r}_1-{\bf r})\delta({\bf r}_n-{\bf r}')
\prod_{i=1}^{n-1} S_{ss}({\bf r}_i, {\bf r}_{i+1})
\label{eq:contractedsegment}
\end{equation}
We then sum the contracted distributions over all possible numbers of like-charged monomers within each segment, 
appropriately weighted by the $NN$ Coulombic penalty function, 
\begin{equation}
{\hat F}^{(s)}({\bf r}, {\bf r}')=  \sum_{n=0}^\infty e^{-\kappa_{++}n}{\bar F}^{(s)}_{n}({\bf r}, {\bf r}')
\label{eq:weightedcontractedsegment}
\end{equation}
These ${\hat F}^{(s)}({\bf r}, {\bf r}')$ are then linked using $S_{+-}({\bf r}, {\bf r}')$ to form a chain of 
$N+1$ segments (with alternating charge) and terminating in a positive segment.  As discussed above these are constructed such that 
odd and even values $N+1$ will have constraints on the charge of the terminal segments.  Below, are the examples $N =0, 1, 2$ and $3$: \newline
\newline
$g_+({\bf r}_1){\hat F}^{(+)}({\bf r}_1, {\bf r}_2)$\newline
$g_-({\bf r}_1){\hat F}^{(-)}({\bf r}_1, {\bf r}_2)S_{-+}({\bf r}_2, {\bf r}_3){\hat F}^{(+)}({\bf r}_3, {\bf r}_4)$\newline
$g_+({\bf r}_1){\hat F}^{(+)}({\bf r}_1, {\bf r}_2)S_{+-}({\bf r}_2, {\bf r}_3){\hat F}^{(-)}({\bf r}_3, {\bf r}_4)S_{-+}({\bf r}_4, {\bf r}_5)
{\hat F}^{(+)}({\bf r}_5, {\bf r}_6)$\newline
$g_-({\bf r}_1){\hat F}^{(-)}({\bf r}_1, {\bf r}_2)S_{-+}({\bf r}_2, {\bf r}_3){\hat F}^{(+)}({\bf r}_3, {\bf r}_4)S_{+-}({\bf r}_4, {\bf r}_5)
{\hat F}^{(-)}({\bf r}_5, {\bf r}_6)S_{-+}({\bf r}_6, {\bf r}_7){\hat F}^{(+)}({\bf r}_7, {\bf r}_8)$\newline
\newline
We will denote these functions as $c^{(+)}_N({\bf r}_1, ....{\bf r}_{2N+2})$.  This function can be contracted as follows,
\begin{equation}
{\bar c}^{(+)}_{N}({\bf r})=  \int d{\bf r}_1... d{\bf r}_{2N+2} \delta({\bf r}_{2N+2}-{\bf r})
c^{(+)}_N({\bf r}_1, ....{\bf r}_{2N+2})
\label{eq:contractdistribution}
\end{equation}
The total positive density charge contribution due to chains of this type (fixed $N_{seg}$) is given by,
\begin{equation}
n^{(+)}_{N_{seg}}({\bf r}) = \phi_P K e^{-\kappa_{+-} (N_{seg}-1)} \sum_{N=1}^{N_{seg}}{\bar c}^{(+)}_{N_{seg}-N}({\bf r}){\bar c}^{(+)}_{N-1}({\bf r})
\label{eq:Nsegden}
\end{equation}
Finally, we can sum this over $N_{seg}$ to obtain,
\begin{equation}
n^{(+)}({\bf r}) = \phi_P K \sum_{N_{seg}=1}^{\infty} e^{-\kappa_{+-} (N_{seg}-1)} \sum_{N=1}^{N_{seg}}{\bar c}^{(+)}_{N_{seg}-N}({\bf r}){\bar c}^{(+)}_{N-1}({\bf r})
\label{eq:positiveden1}
\end{equation}
which can be rewritten as,
\begin{equation}
n^{(+)}({\bf r}) = \phi_P K {\hat c}^{(+)}({\bf r}) {\hat c}^{(+)}({\bf r})
\label{eq:positiveden2}
\end{equation}
\begin{equation}
{\hat c}^{(+)}({\bf r}) =  \sum_{N_{seg}=1}^{\infty} e^{-\kappa_{+-} N_{seg}} {\bar c}^{(+)}_{N_{seg}}({\bf r})
\label{eq:positiveden1}
\end{equation}
Similarly, we obtain for negative charges,
\begin{equation}
n^{(-)}({\bf r}) = \phi_P K {\hat c}^{(-)}({\bf r}) {\hat c}^{(-)}({\bf r})
\label{eq:positiveden2}
\end{equation}
We can obtain the following recursion formulae for $\hat{c}^{(+)}({\bf r})$ and $\hat{c}^{(-)}({\bf r})$.
\begin{eqnarray}
\hat{c}^{(+)}({\bf r}) & = &  g_+({\bf r}) + e^{-\kappa_{+-}} S_{+-}({\bf r},{\bf r}') \ast \hat{c}^{(-)}({\bf r}')  
+ e^{-\kappa_{++}} S_{++}({\bf r},{\bf r}') \ast \hat{c}^{(+)}({\bf r}')\\
\hat{c}^{(-)}({\bf r}) & = &  g_-({\bf r}) + e^{-\kappa_{+-}} S_{+-}({\bf r},{\bf r}') \ast \hat{c}^{(+)}({\bf r}')  
+ e^{-\kappa_{--}} S_{--}({\bf r},{\bf r}') \ast \hat{c}^{(-)}({\bf r}') 
\label{eq:recursion}
\end{eqnarray}
with the asterix defining a convolution.
The above formulae then form a set of self-consistent equations to be solved for the overall densities.

\subsection{Interactions between charged surfaces, in the presence of dissociated $1:1$ salt}
We will explore interactions between charged surfaces immersed in an aqueous solution
containing a monovalent ($1:1$) salt, under the assumption that a fraction
of the simple ions bond to form ion clusters. The latter category
of ions, which we denote as ``monomers'', follow
the cluster formalism described above, whereas the non-clustering part
remain as ``dissociated" ions. With this distinction, the grand potential, Eq. (\ref{eq:grandpot})
needs to be modified somewhat, to include the corresponding free energies and
chemical potentials of the dissociated ions. In particular, the excess free energy $F^{ex}$ is now
a function of monomer as well as dissociated ion densities. This function can in turn be
decomposed into a term that describes Coulomb interactions between all charged
species (including wall charges), and another term that accounts for excluded volume interactions. 
The excluded volume free energy cost is approximated by a Carnahan-Starling
expression, where all particles carry a hard sphere of diameter $d_{hs}=3$ {\AA}, but with the
monomer excluded volume reduced by a factor, due to overlapping of free 
volume between adjacently bonded monomers. This is a similar approximation to that used
in an earlier study \cite{Woodward:92}. We emphasise that the conclusions
made in the current work is not sensitive to the way in which excluded
volume effects are approximated.

We recall that monomers are characterised by the common notation ``$s$'', the sign of
which identifies the monomer valency. We shall, in a similar fashion, use ``$d$'' to
characterise dissociated (non-clustering) ions, where again the
sign identifies the valency of those ions ($d$ should not be confused with $d_{hs}$). 

The final term
in Eq. (\ref{eq:lambda}) can be split into two separate contributions:
\begin{equation}
  \psi_s^0({\bf r}) = V_{wall}({\bf r})+se\psi_s^{D_s}({\bf r})
\label{eq:psipot} 
\end{equation}
where the first term describes the non-electrostatic component of the surface potential. The two walls
are assumed to be parallel, with an infinite extension along the $(x,y)$ coordinates,
and located at $z=0$ and $z=h$, respectively. Thus $V_{wall}$ is zero in the
bulk, as well as in the region $d_{hs}/2<z<(h-d_{hs}/2)$ and infinite elsewhere.
The last term describes the monomer valency ($s$) multiplying the elementary charge ($e$) times
a constraining potential, $\psi_s^{D_s}({\bf r})$, which is zero in the bulk solution and equal to the Donnan
potential, $\psi^{Donnan}$, inside the slit, for systems at full electro-chemical equilibrium. However, we will also explore the
option of a ``semi-restricted'' equilibrium.  In this case we will assume that the surface charge
is neutralised only by the dissociated ions, while the total charge of monomers, summed over all 
the ionic clusters, is assumed to be zero.  The reasons for this latter assumption
is argued below. 

The corresponding expression to Eq.(\ref{eq:psipot}) for dissociated ions reads:
\begin{equation}
  \psi_d^0({\bf r}) = V_{wall}({\bf r})+de\psi_d^{D_d}({\bf r})
\label{eq:psipotd} 
\end{equation}
where we recall that $d$ is the valency of the dissociated charges.
It should be noted that $\psi_d^{D_d}({\bf r})$ also acts on the wall
charges. At full equilibrium, $\psi_d^{D_d}({\bf r})$ is equal to the
Donnan potential, for all species inside the slit.
At semi-restricted equilibrium, where the surface charge is fully neutralised
by dissociated ions, the potential is slightly different, as we shall
demonstrate below. All calculations are made at room temperature, with a Bjerrum length of $7.16$ {\AA}.
The bond length between connected monomer is set to $b=4$ {\AA}. 

Defining $g_s(h) \equiv \Omega_{eq}/A-p_bh$, where $p_b$ is the bulk pressure, we can
write the {\em net} interaction free energy as $\Delta g_s = g_s(h)-g_s(h_{max})$, where
$h_{max}$ is at large separation so that the net interaction can be assumed to be zero.
The net pressure acting perpendicularly to the flat
surfaces is denoted $p_{net}$, with $p_{net} = -\partial g_s/\partial h$. This quantity
can be evaluated either analytically or discretely (from $g_s(h)$) and the agreement
between these forms an important and useful check of the cDFT codes.

\section{Results}
While most of our efforts will be devoted to models of ion clusters (with charged monomers) it is instructive 
to first probe the qualitative responses to parameter changes of 
toy $A/B$ mixture model, composed of simple uncharged hard-sphere monomers, labelled $A$ and $B$.

\subsection{Toy model.}
Here, the $A$ and $B$ particles are simple hard spheres, of the same diameter: $d=3$ {\AA}. 
They are only distinguishable by their affinity to the surfaces. Both monomer types are excluded from
the internal surface region by $V_{wall}$.
However, the $A$ monomers also experience
an adsorption potential $w$, which at the left wall can be written as:
\begin{equation}
w(z) =   \left\{
	\begin{array}{ll}
		0 ; & z \geq z_c \\
		C((z-d/2)^2-(z_c-d/2)^2)/d^2 & z < z_c  \\
	\end{array}
        \right.
\end{equation}
where we have set $z_c=2d$ and $\beta C=0.45$. The total affinity between an $A$ monomer
and the surfaces separated by $h$ is $W_{wall}(h,z)=w(z)+w(h-z)$.
For clarity we will replace the $\kappa_{++}$ and $\kappa_{+-}$ notation with $\kappa_{AA}$ and $\kappa_{AB}$
in the toy model. It should be noted that in our $A/B$ mixture, there is no constraint on the
sign of $\gamma$, in contrast to ion clusters, where we expect oppositely
charged monomers to be bonded more tightly on account of the extra Coulomb attraction.

\begin{figure}[h!]
	\centering
	\includegraphics[scale=1]{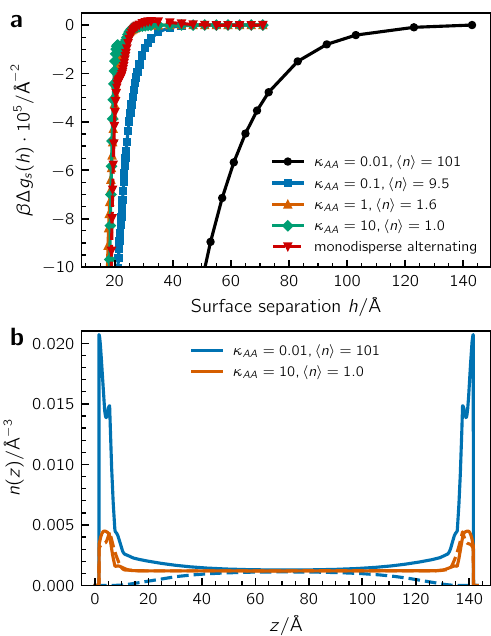}
	\caption{\textbf{a} $A/B$ mixtures: net interaction free energies. \textbf{b} $A/B$ mixtures: density profiles. Solid and dashed lines display distributions
	      of $A$ and $B$ monomers.}
	 \label{fig:AB}
\end{figure}
  
The resulting surface interactions and monomer density profiles are
displayed in Figure \ref{fig:AB}. The average polymer length is held fixed at $\langle r \rangle\approx 101$ but
the degree of ``polarisation'' as determined by the $bulk$ parameter, $\langle n \rangle$, is varied. 
A large value of $\kappa_{AA}$ results in
chains with predominately an alternating sequence ($ABAB...$).  A value of $\kappa_{AA}=10$ is high enough to maintain
this alternating sequence even in the presence of surfaces attractive to A. The corresponding
surface interaction (graph \textbf{a}) is rather short-ranged and monotonically attractive, due
primarily to bridging attractions.  For adsorbing chains one may expect an intermediate
barrier, as surfaces approach each other and adsorbed chains first encounter each other's excluded
volume.    This barrier is present in monodisperse chains but is absent in 
 adsorbed ``living'' chains, as described in the current model \cite{Woodward:08a}.
Living chains are able to alleviate hard core repulsions by suitably varying their length.
By decreasing $\kappa_{AA}$ we are able to induce polarization in the chains, but it is only for values
significantly below $1$ that we observe a strong effect on the surface forces. Nevertheless, at
$\kappa_{AA}=0.1$, the range of the attraction has increased somewhat, and
$\kappa_{AA}=0.01$ leads to a very long-ranged attractive tail. Note that the latter
value effectively generates (polydisperse) homopolymers.

The density profiles in  Figure \ref{fig:AB}(b), show that a decrease in  $\kappa_{AA}$ causes
a much stronger adsorption at the walls. In contrast, with an alternating sequence (large $\kappa_{AA}$), the
required presence of non-adsorbing $B$ effectively reduces the overall surface affinity.

\subsection{Ion clusters}
We now turn our attention to ion clustering, and the influence of these clusters on interactions between charged surfaces. 
In earlier work, we have shown that, if a {\em dominant} fraction of
ions form largely neutral (or at most singly charged) and monodisperse clusters, then
quite strong and long-ranged interactions are produced \cite{Ribar:25a,Ribar:25c}.
One could argue that significant clustering is not consistent with conductivity measurements. While there is a
tendency for the (linear) dependence of conductivity on concentration to decrease
at high ionic strengths, the effect is rather modest, at least for $NaCl(aq)$ solutions\cite{Kamcev:18}.

Given these arguments,  we will here consider a rather different scenario to what we did previously.
More specifically, we will consider here cases where only a {\em minor} fraction of the ions form clusters
of the {\em polydisperse} type described in the present model.
We begin by solving the cDFT, as is, assuming  so called {\em full electro-chemical equilibrium} for all the 
charged species.  

\subsubsection{Full equilibrium}
\begin{figure}
	\centering
	\includegraphics[scale=1]{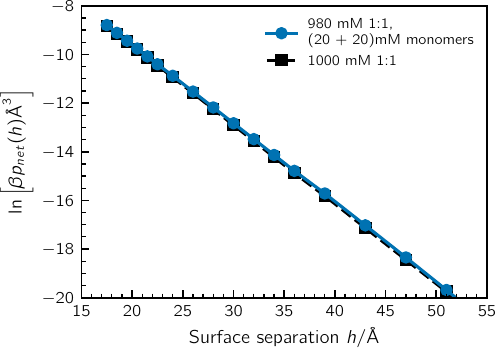}
	
	\caption{A comparison between the long-ranged net pressure tails
		at approximately 1 M dissociated salt, in the absence and presence of
		20 mM ``monomer salt''. The latter ion type can polymerise as well as polarise:
		$\langle r \rangle=25$ and $\langle n \rangle\approx 1.14$. While not shown here, the results
		are very similar with fully alternating chains, $\langle n \rangle=1.00$.}
	
	\label{fig:g_fulleq}
\end{figure}

For full equilibrium, we have that the electrochemical equilibria both 
associated (monomers) and dissociated ions give rise
identical  Donnan potentials, i.e., $\psi_s^{D_s} = \psi_d^{D_d} = \psi^{Donnan}$.
In this case, the impact from ion clusters on surface interactions is quite
modest. This is shown in Figure \ref{fig:g_fulleq} for 
an ion mixture of 1 M dissociated electrolyte, in the absence and presence of
20 mM of "monomer salt''. The latter ions are characterised by an average bulk polymer length of
$\langle r \rangle=25$ and and average like-charged segment length of $\langle n \rangle\approx 1.14$.
Thus there are significantly more dissociated than associated ions and the degree of polarisation 
allowed for the clusters is quite low, as $\langle n \rangle$ is close to unity.  
The surface affinity is much greater for dissociated ions than ion clusters (polyampholytes). 
The reason is that the net charge of the polyampholytes typically speaking is quite modest (close to zero) leaving clusters
depleted from the near-surface region and having very little impact 
on the surface forces.  
\subsubsection{Semi-restricted equilibrium}
In the cDFT as formulated here, the propensity to form ion clusters is determined by nearest-neighbour 
parameters $\kappa_{+-}$ and $ \kappa_{++}$.  These parameters in essence manifest the 
intrinsic free energy of the clusters (albeit in the bulk fluid).  Unfortunately, the model contains no 
further considerations of the internal correlations within the clusters, which will likely be very strong,
considering the local dielectric constant is diminished due to the lack of freely rotating water molecules 
within.  These internal thermodynamic considerations will be non-local, i.e.,  they will
act beyond nearest-neighbours.  A plausible outcome of this is that the accumulation of charge
on some part of a given cluster will be balanced to some degree by a counter charge within that cluster
and in adjacent, neighbouring clusters.  This assumes that inter- and intra-cluster correlations
are stronger than those involving dissociated ions.   For this reason, it is plausible to impose 
a constraint on the solutions of the cDFT, which reflects the different correlations between monomer ions within
clusters and those among dissociated ions.  

A simple way to realise these differences is applied here.  Specifically, we will make the 
assumption that the neutralisation of the charged surfaces is carried out by the dissociated ions,
while the charge of the clustered (monomer) ions within the region between the surfaces sums to zero.  
This allows for charge polarisation of the clusters, but insists that the clusters (due to their stronger correlations) must essentially
neutralise themselves and/or each other.  This global constraint can be viewed as one of the simplest
examples of a potential family of cluster constraints that reflect the different responses of bound and dissociated ions. 
In this context, we can also mention that the general dynamics of clusters are 
expected to be somewhat slower than dissociated ions, including the rearrangement of ions
within clusters.  This has often been used as an argument to impose a semi-restricted equilibrium 
constraint on cluster response.   

\begin{figure}
  \centering
    	\includegraphics[scale=1]{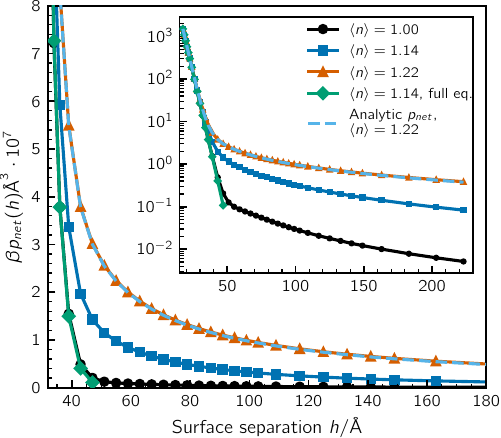}
    
        \caption{The net pressures, at semi-restricted equilibrium, with $\langle r \rangle=25$ and various
          degrees of polarisability, $\langle n \rangle$. The bulk solution contains $980$ mM dissociated
          salt and $20+20$ mM monomers (cat- and anionic).  
          The inset displays the long-range decay, as
          illustrated by the logarithm of the net pressure.
          The dotted line, for $\langle n \rangle=1.22$, demonstrates the equivalence between
          the analytic ($p_{net} = -\partial g_s/\partial h$) and
          discrete ($p_{net} = -\delta g_s/\delta h$) free energy derivatives.
          The corresponding net pressure at full
          equilibrium is shown for reference, by a green line with diamonds.
    }
	\label{fig:p_vn}
\end{figure}

Pragmatically, this constraint is carried out in the cDFT by allowing different values for $\psi_s^{D_s}$  and $ \psi_d^{D_d}$.    
The consequences are rather dramatic.  As can be seen in Figure \ref{fig:p_vn}, the slight deviation from complete
equilibrium afforded by this constraint, generates a long-ranged repulsive
force, even with a minor fraction of clustering species. This is the same system displayed in 
Figure \ref{fig:g_fulleq} at full equilibrium. The interactions at
full and semi-restricted equilibrium agree quite well at short separations, but
deviate dramatically at intermediate and large separations. 

One of the most peculiar aspects of ``anomalous underscreening'' is the way in which the
interactions between charged surfaces displays a range that
either is roughly constant, or even {\em increases} as more
salt is added, beyond concentrations of 1 M, or thereabout.
Let us explore predictions in this regime of the semi-restricted equilibrium constraint hypothesis.

\begin{figure}
	\centering
	\includegraphics[scale=1]{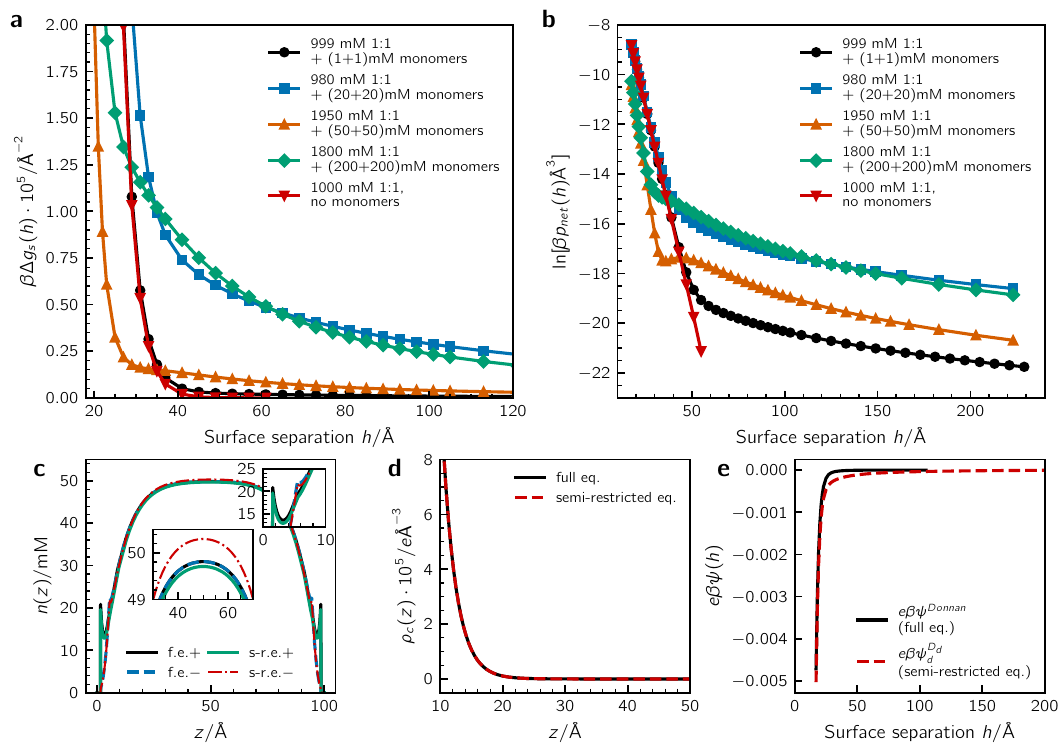}
	\caption{\textbf{a} Illustrations of possible
		      effects on surface interactions, obtained by adding salt,
		      under the assumption of semi-equilibrium conditions
	      and concentration-induced clustering. Bond parameters ($\kappa_{++}$ and $\kappa_{+-}$)
	      are chosen such that $\langle r \rangle=25$ and $\langle n \rangle \approx 1.14$ in all
              cases. \textbf{b} Long-range $p_{net}$ decay. \textbf{c} Comparing density ($n(z)$) and \textbf{d} charge density ($\rho(z)$) 
		      profiles at a surface separation
		      of $100$ {\AA}. The dashed lines in graph \textbf{c} show anion profiles. The bulk concentration of monomers
		      and dissociated salt is $50$ mM and $1950$ mM (commensurate with a
		      $2$ M salt solution, $50$ mM of which form clusters).
		      Results are shown from calculations at full and
		      semi-restricted equilibrium. \textbf{e} Comparing the variation of restriction potentials $\psi^{Donnan}$ (full equilibrium)
		      and $\psi_d^{D_d}$ (semi-restricted equilibrium)
		      with separation. The bulk concentrations of monomers
		      and dissociated salt are $50$ mM and $1950$ mM, respectively.}
		      \label{fig:semi_eq}
\end{figure}

We  first imagine a salt solution of 1 M, in which some small fraction
are cluster-forming ions (monomers). Upon an increase of the overall
concentration to 2 M, one would anticipate that the fraction of clustering
ions increases, though the $specific$ molecular mechanism underlying the
cluster formation is beyond the scope of this work. We examine
possible scenarios, given these preliminary considerations.
In Figure \ref{fig:semi_eq}, we have plotted interaction free energies (graph \textbf{a}) and
net pressures (graph \textbf{b}) at an overall total salt concentration of 1 M and 2 M, where
we make the qualitative assumption that a smaller fraction of ions cluster
at the lower concentration. We see that if this doubling of the overall
salt concentration leads to a ten-fold increase of the
clustering ion (monomer) concentration (say $20$ mM $\Rightarrow$ $200$ mM)
then the surface forces remain similar.
But if the cluster tendency grows more rapidly with ionic strength
(say $1$ mM $\Rightarrow$ $200$ mM), then we find a substantially increased strength at the higher
concentration, although the range  appears to be roughly
constant.

The origin of the strong effect that this seemingly modest deviation from full equilibrium can be sought as follows. 
In Figure \ref{fig:semi_eq}\textbf{c} and \textbf{d}, we compare monomer and charge density profiles
at a surface separation of $100$ {\AA}, under full and
semi-restricted equilibrium, in a solution containing $1950$ mM dissociated
ions, and $(50+50)$ mM cat- and anionic monomers. The depleted
monomer profiles that we discussed earlier, are explicitly shown in
graph (\textbf{c}), where we also note a strong similarity between profiles
established at full and semi-restricted equilibrium. Nevertheless,
there is a small but detectable (inset) difference that progresses all the way to the
mid-plane of the slit. On the other hand, graph (\textbf{d}) illustrates that
the corresponding {\em charge density} profiles are essentially identical. 
This in turn implies an almost identical local potential variation across the slit. 

%
In Figure \ref{fig:semi_eq}\textbf{e} we see that the strength of the constraining potential
$\psi_d^{D_d}$ decays much slower than its
correspondence at full equilibrium, $\psi^{Donnan}$. The deviation
is rather small in absolute terms. Nevertheless, this pinpoints
the origin of the large differences we have seen between
surface interactions at full and semi-restricted equilibrium.

Finally we consider the effect of cluster size, i.e., average polymer length. 
Our two-parameter description of
ion-ion bonds, suggests that there is more than one
way to adjust our parameters in order to change the
average degree of polymerisation, $\langle r \rangle$ - cf. Figure \ref{fig:kappa_dep}.
\begin{figure}
	\centering
	\includegraphics[scale=1]{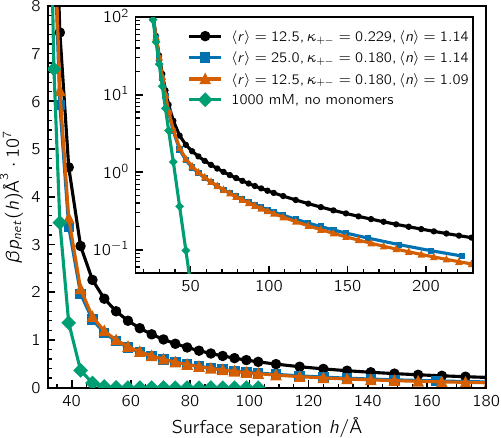}
	\caption{Examples of how the surface interactions can vary with
		the average cluster size, under the assumption of
		semi-restricted equilibrium. In the cluster-containing solutions, the monomer
		concentration is $(20+20)$mM and the concentration of dissociated salt is $980$ mM. The insert presents the
                long-ranged net pressure decays.}
	\label{fig:avrdep}
\end{figure}
In Figure \ref{fig:avrdep} we show examples of how the 
surface forces can change, at semi-restricted equilibrium, if the
average polymer length is reduced. We note, perhaps somewhat surprisingly, that
the repulsion might become stronger as the average cluster size drops.
We should then note  that this reduction is accompanied by
an overall increase of the cluster concentration, since the density
of the cluster-forming monomers is constant. Another aspect to keep in mind
is that for very small average cluster sizes, the underlying mechanisms
that (might) support semi-restricted conditions become weaker, i.e. for
tiny average cluster sizes, there is a smaller difference in terms of
charge density, when compared with fully dissociated ions.

\section{Conclusions}
We have developed a cDFT formalism to treat mixtures of simple $A$ and
$B$ particles, that can reversibly associate by breaking or
forming $A-A$, $A-B$ and $B-B$ bonds . While the cDFT treatment is general, we have
devoted particular attention to its application on a
hypothetical scenario in which simple monovalent ions
cluster together by forming reversible bonds at high salt concentrations in aqueous solutions.
A similar scenario could appear in ionic liquids, or ionic liquid+solvent mixtures, with
an analogous cluster formation mechanism. We have not investigated what
underlying molecular mechanisms might generate the reversible bonds in this work.

Assuming that only a rather small fraction of the total
amount of ions polymerise to form (widely polydisperse) linear clusters, our mean-field calculations
indicate that the presence of these clusters only have a minor
influence on the interaction between charged surfaces, at complete equilibrium.
We note that dissociated ions, with a much higher concentration and surface charge density than a typical cluster, accumulate
at the surfaces, and displace the clusters so that the latter are depleted in these regions.

The expected strong correlations within and between clusters, perhaps
combined with an expected slow cluster dynamics suggests a simple 
semi-constrained model  that surface neutralisation is entirely accomplished by dissociated ions.
This raises a more general question as to what extent a concentrated ionic solution  
necessarily maintains complete equilibrium during a typical SFA experiment.
We have explored this semi-restricted model and found that only a very
small fraction of cluster-forming ions suffices to generate
strong and long-ranged surface forces. We speculate here that
this phenomenon may be related to the observed ``anomalous underscreening''
though further investigations on other types of constrained systems are
clearly required to draw firmer conclusions. 
It is possible that the constraint used here 
exaggerates the differences in cluster and dissociated ion
correlations.   In ``real'' solutions one might
anticipate that deviations from complete equilibrium primarily is
relevant for rather large clusters. Still, we believe that our approach is a  first step 
in establishing the use of  plausible constraints in simple theories
(such as the mean-field cDFT) to investigate these phenomena.
We find it particularly intriguing that
even though the semi-equilibrated restriction has a 
very small influence ion distributions, and a
vanishing impact on charge density distributions, it nevertheless
produces quite dramatic changes to the surface forces, also when
the total cluster concentration is quite low. 
This type of generalised cDFT formalism could also prove useful to
a broader range of systems than that investigated in this work. 

\begin{acknowledgement}
  J.F. acknowledges financial support by the Swedish Research Council.

\end{acknowledgement}




\bibliography{poly.bib}

\end{document}